\documentstyle[aps,prd]{revtex}

\begin{document}
\draft
\title{First results for the Coulomb gauge integrals using NDIM}
\author{Alfredo T. Suzuki\footnote{e-mail:suzuki@ift.unesp.br} and Alexandre G. M.
Schmidt\footnote{e-mail:schmidt@ift.unesp.br}}\address{ Instituto de
F\'{\i}sica Te\'orica, Universidade Estadual Paulista, R.Pamplona, 145 
S\~ao Paulo - SP CEP 01405-900 Brazil}
\date{\today}
\maketitle

\begin{abstract}
The Coulomb gauge has at least two advantadges over other gauge
choices in that bound states between quarks and studies of confinement
are easier to understand in this gauge. However, perturbative
calculations, namely Feynman loop integrations are not well-defined
(there are the so-called energy integrals) even within the context of
dimensional regularization. Leibbrandt and Williams proposed a
possible cure to such a problem by splitting the space-time dimension
into $D=\omega+\rho$, i.e., introducing a specific one parameter $\rho$
to regulate the energy integrals. The aim of our work is to apply
negative dimensional integration method (NDIM) to the Coulomb gauge
integrals using the recipe of split-dimension parameters and present
complete results -- finite and divergent parts -- to the one and two-loop
level for arbitrary exponents of propagators and dimension.

\end{abstract}
\vspace{.5cm}

\def\be{\begin{equation}}
\def\ee{\end{equation}}
\def\beq{\begin{eqnarray}}
\def\eeq{\end{eqnarray}} 
\def\s{\sigma} 
\def\p{\rho}
\def\pp{{\bf p}}
\def\G{\Gamma} 
\def\F{_2F_1} 
\def\an{analytic} 
\def\ac{\an{} continuation} 
\def\hsr{hypergeometric series representations} 
\def\hf{hypergeometric function} 
\def\ndim{NDIM} 
\def\quarto{\frac{1}{4}} 
\def\half{\frac{1}{2}} 
\section{Introduction}

Perturbative approach in quantum field theory (QFT) was responsible
for several breakthrough ideas in Physics and Mathematics. One of them
is dimensional regularization\cite{bolini}, i.e., \ac{} of space-time
dimension $D$ into an extended domain that allows for complex
values. Feynman loop integrals gained a solid theoretical foundation
and renormalization process became simpler than it was (where one had
to use cut-offs and so on). Of course, this is only a partial picture
of it all, say, the covariant side of the coin.

In algebraic non-covariant gauges\cite{leib}, on the other hand,
dimensional regularization was able to control divergences, e.g., in
the light-cone gauge, but the results were not physically
acceptable. In other words, double-poles did appear in one-loop
integral calculations and Wilson loops did not have the correct
behavior\cite{caraciolo}. These problems were first overcome with the
advent of what is known as the Mandelstam-Leibbrandt (ML)
prescription\cite{mandelstam}. More recently, we have shown that in
the NDIM approach we do not need to invoke any kind of prescription to
perform Feynman loop integrals in this gauge\cite{prescless}.

Among the non-covariant gauges we also have the Coulomb gauge (often
referred to as the radiation gauge) where confinament\cite{zwanziger}
and bound states\cite{kummer} are easier to deal with, ghost
propagator has no pole and unitarity is manifest. However, in such a
gauge, no further insight has beeen achieved with the standard
dimensional regularization technique, because it presents a gauge
boson propagator of the form

\be 
G_{\mu\nu}^{ab} (q) = -\frac{i\delta^{ab}}{q^2}\left[ 
\eta_{\mu\nu} +\frac{ n^2}{{\bf q}^2} q_\mu q_\nu - \frac{q\cdot n}{{\bf 
q}^2} \left( q_\mu n_\nu+ n_\mu q_\nu \right)\right], \qquad{\rm
with }\,\, n_\mu=(1,0,0,0), 
\ee 
which generates loop integrals like,

\be 
\int \frac{d^Dq}{q^2 ({\bf q-p})^2}. 
\ee
where bold face letters stand for three momentum vectors.

The integral over the fourth-component (in Euclidean space) or
zeroth-component (in Minkowski's), the so-called energy integral, is
not defined even within the context of dimensional
regularization. Doust and Taylor\cite{taylor} discussed Coulomb gauge
loop integrals and presented a possible remedy for this problem in
terms of a interpolating gauge (between Feynman and Coulomb, see also
\cite{halpern}). Leibbrandt\cite{split}, again, and Williams,
presented another approach for this ill-defined integrals, a procedure
called {\it split dimensional regularization}. Both parts, namely
energy and 3-momentum sectors, need to be separately dimensionaly
regularized, that is, one parameter only, $D$, is not sufficient to
render the integrals well-defined. To overcome this problem, they
introduced another regulating parameter, i.e., split the dimensinality
of space-time into two distinct sectors, namely,
$D=4-2\epsilon=\omega+\rho$ and the divergences contained in energy
integrals are expressed as poles in $\rho$ besides the usual ones in
terms of $\omega$, that is to say, the integration measure is written
down as $d^Dq=d^\omega {\bf q}\; d^\rho q_4$.

In a series of papers\cite{leib-coulomb}, Leibbrandt studied Coulomb 
gauge integrals to one and two-loop level (with Heinrich) and presented 
results for divergent parts of several of them. Our aim, in this work is 
two-fold: show that NDIM is the most versatile technique to 
carry out loop integrals, whether they come from covariant or non-covariant 
gauges; and to present complete results for the Coulomb gauge integrals to 
one and two-loop level for arbitrary exponents of propagators and 
dimension.

The outline for our paper is as follows: in section II we consider scalar 
and tensorial Coulomb gauge integrals at one-loop level, while section III is 
devoted to two-loop integrals and in section IV we present our 
conclusions. In the appendix we discuss some technical issues.

\section{One-loop Coulomb gauge integrals}

To show how NDIM can handle Coulomb gauge integrals with ease we
consider in this section one-loop integrals. Recall that negative
dimensional integration is equivalent to positive dimensional
integration over Grassmannian variables\cite{halliday} --- a property
demonstrated by Dunne and Halliday --- and for this very reason,
propagators are raised to positive powers (they appear in the
numerator of integrands) and usual variables become Grassmannian
ones. Another important point is that in the NDIM context it is as
simple to work with arbitrary exponents of propagators as if we choose
particular values for them. This is why we consider the general
case. It is also worth remembering that for some types of diagrams,
e.g., box integrals\cite{box,stand}, there are divergences that are
not related to space-time dimension $D$, but they appear as poles for
particular values of exponents, say $i$ and $j$, of propagators,
yielding singularities expressed as, for instance $\G(i-j)$, see also
\cite{ussyukina}. So, within the NDIM approach we can also trace back
the origin of divergences.

The first two integrals we choose to work with are scalar ones,

\be 
g_1(i,j,k) = \int d^D\! q\; (q^2)^i (q+p)^{2j}\; ({\bf q}^2)^k ,
\ee

\be 
g_2(i,j,k) = \int d^D\! q\; (q^2)^i ({\bf q+p})^{2j}\; ({\bf q}^2)^k ,
\ee
where generating functions for these referred Coulomb gauge integrals are,

\beq
\label{int1}
G_1 &=& \int d^D\! q\; \exp{\left[ -\alpha q^2 - \beta (q+p)^2 - 
\gamma {\bf q}^2 \right]}, \\
G_2 &=& \int d^D\! q\; \exp{\left[ -\alpha q^2 - \beta {\bf (q+p)}^2 - 
\gamma {\bf q}^2 \right]}, 
\eeq
with $D=\omega+\rho=4-\epsilon\;\;$ and $d^D\! q=d^\omega{\bf q}\;
d^\rho q_4$, in Euclidean space, following the split dimension recipe
of Leibbrandt {\em et al}.

Completing the square we can easily carry the integration out to get,

\beq 
\label{int2}
G_1 &= & \left(\frac{\pi}{\alpha+\beta} \right)^{\rho/2}
\left(\frac{\pi}{\lambda_1} \right)^{\omega/2} \exp{\left[ -\frac{
(\alpha+\gamma)\beta {\bf p}^2}{\lambda_1}\right] } \exp{ \left( -
\frac{\alpha\beta p_4^2}{\alpha+\beta}\right) }, \\
G_2 &= & \left(\frac{\pi}{\alpha} \right)^{\rho/2} \left(\frac{\pi}{\lambda_1}
\right)^{\omega/2} \exp{\left[ -\frac{ (\alpha+\gamma)\beta {\bf 
p}^2}{\lambda_1}\right] }, 
\eeq
where $\lambda_1= \alpha+\beta+\gamma$.

Taylor expanding both expressions in (\ref{int1}) and (\ref{int2})
we get the NDIM solutions for $g_1$ and $g_2$ by solving systems of
linear algebraic equations.

The system of linear algebraic equations for the first integral is given 
by a $5\times 8$ matrix, 
$$ \left\{ \matrix{ X_{13} + Y_{14} = i \cr
                     X_{123} + Y_{25} = j \cr
                     X_2 + Y_3 = k \cr
                     Y_{123} = -X_{12}-\omega/2 \cr
                     Y_{45} = -X_3 - \p/2 }\right., $$ 
with five equations and eight ``unknowns'', corresponding to the
various summation indices coming from the Taylor and multinomial
expansions. They are solvable only within the lower quadratic $5\times
5$ dimension matrices. There are a grand total of 56 possible square
matrices of this type (i.e., $5\times 5$) from which 36 yield relevant
non vanishing and workable solutions while the remaining 20 yield a
set of trivial solutions (i.e., the related systems do not have a
solution). We know from our previous works (see for instance
\cite{box}) that all those non-trivial solutions will generate power
series of hypergeometric type, known as \hf{}s\cite{luke}. Moreover,
all of them are related by \ac{}, either directly or indirectly. In
our present case, namely integral $g_1$, there are triple as well as
double series, among which we choose to consider only the simplest
ones,

\be 
g_1^{A\;[{\rm AC}]} (i,j,k) = f_1^{A,\;[{\rm AC}]} \sum_{n_1,n_2,n_3=
0}^\infty \left( \frac{\pp ^2}{p_4^2}\right)^{n_{123}} \frac
{ (-1)^{n_2}(-i|n_{123}) (k+\omega/2|n_{12}) (D/2+j|n_{23})
(1-i-\p/2|n_{123}) }{n_1!n_2!n_3!  (1+j+k+D/2|n_{23})
(1-i-\rho/2|n_{12}) (\s+\omega/2-i|n_{123})} , 
\ee
where the superscript ``[{\rm AC}]'' means analytic continuation (to positive
dimensional region) and we define the shorthand notation $n_{AB} = n_A
+ n_B$, while $(x|y) \equiv (x)_y = \G(x+y)/\G(x)$ is the Pochhammer symbol
and
 
\be 
f_1^{A,\;[{\rm AC}]} =  \pi^{D/2} (-p_4^2)^i(\pp ^2)^{j+k+D/2} 
(-j|j+k+\omega/2) (-k|j+k+D/2) (-i+\s+\omega/2|i-\s-\omega/2-D/2-j-k),
\ee
where $\sigma=i+j+k+D/2$. Observe that the above result is valid for
negative $j,k$. Among the 36 possible series this is the only one that
has the form $\Sigma (\cdots)^{a+b+c}$, where the ${\cdots}$ stands for
the specific kinematical configuration.

For the case of double series we have, e.g.,

\be 
g_1^{B,\;[{\rm AC}]} (i,j,k) = f_1^{B,\;[{\rm AC}]} \sum_{n_1,n_2=
0}^\infty \left( \frac{\pp ^2}{p_4^2}\right)^{n_{12}} \frac
{ (-\s|n_{12}) (-k|n_2) (\omega/2+k|n_1) (1-k-\s-D/2|n_2) }{n_1!n_2!
(\omega/2|n_{12}) (1-j-k-D/2|n_2)} ,
\ee
where 
\be 
f_1^{B,\;[{\rm  AC}]} =  \pi^{D/2} (p_4^2)^\s (-i|\s) (-j|\s) (\omega/2|k) 
(k+\s+D/2|-2\s-k-D/2) ,
\ee
and the result is valid when $i,j$ are negative.

This hypergeometric series representation is 4-fold degenerate. Here
we mention an important point in the process of \ac{} to positive
dimension and negative values of exponents of propagators referred to
above. The result for the negative dimensional space region for
$g_1^B$ is in fact given by a sum of two terms -- the second one being
also 4-fold degenerate -- however, when we perform the \ac{} this
second term vanishes because it contains a factor of the form
(forgetting about the minus sign)

$$ \frac {1}{(1|-\p/2)} = \frac{\G(1)}{\G(1-\p/2)} \longrightarrow
^{\!\!\!\!\!\!\!\!\!\!\!  \! AC} \;\; (0|\p/2) =
\frac{\G(\p/2)}{\G(0)}=0,
$$
which always vanishes (see also \cite{oleari}) since $\p \neq 0$ by
definition and where, as usual in \ndim{} approach we make use of the
Pochhammer's symbol property $(a|b)=(-1)^b/(1-a|-b)$.

To close this part of the computation, we just mention that of course
there are other hypergeometric series which represent the same Feynman
integral in other kinematical regions, e.g., there is a double series
of the form,
$$
\sum_{a,b=0}^\infty \frac{\G(...)}{a!b!} \left(\frac{p_4^2}{\pp ^2}
\right)^a, 
$$
that is, one of the series (with summation index $b$) has unit
argument and can be recast as a $\,_3F_2(...|1)$. 

These are just the few different manners in which we may write down
the result for the integral $g_1$.

\vspace{.5cm}

The second integral is easier than the first, and its result is also
degenerate, i.e., there are a total of five $4\times 4$ systems to be
solved of which one has no solution and the remaining four, after
properly summed give the same result, yielding

\be 
g_2 =  (-\pi)^{D/2} ({\bf 
p}^2)^\s\frac{\G(1+i)\G(1+j)\G(1-\s-\omega/2)
\G(1+i+k+\rho/2) }{\G(1+\s) \G(1+i+\rho/2) \G(1-i-k-D/2) 
\G(1-j-\omega/2)}, 
\ee
which after analytically continuing to positive $D$ becomes,
\be 
g_2^{[{\rm AC}]}(i,j,k) =  \pi^{D/2} ({\bf p}^2)^\s (\s+\omega/2|-2\s-\omega/2)
(-i|-\rho/2) (-j|\s) (-i-k-\rho/2|\s). 
\ee

\vspace{.5cm}

Next we consider Feynman integrals in the Coulomb gauge with tensorial
structures. Again, as we treated in \cite{tensor} the case of
covariant gauge we show here how NDIM can handle these Coulomb gauge
tensorial integrals in a similar manner. Let,

\be
\label{int-g3} 
g_3(i,j,k) =  \int d^D\! q\; (q^2)^i ({\bf q+p})^{2j} (2{\bf q\cdot
p})^k  ,
\ee
and
\be
\label{int-g4} 
g_4(i,j,k,m) =  \int d^D\! q\; (q^2)^i ({\bf q+p})^{2j} ({\bf q}^2)^k
({2\bf  q\cdot p})^m, 
\ee
so that, after some algebraic manipulations, we eventually get the result,
\be 
\label{g3} g_3^{[{\rm AC}]}(i,j,k) =  \pi^{D/2} (-2)^k(\pp^2)^\s (-i|-j-D/2)
(\s+\omega/2|j-\s)  (-j|\s)\; _3F_2(\{3\}|1) ,
\ee
where the set of parameters $\{3\}\equiv \{a_3,b_3,c_3;e_3,f_3\}$ for
the hypergeometric function ${}_3F_2(\{3\})$ is given in the table.

We must observe here that for (\ref{int-g3}) the exponent $k \geq 0$
always, and in (\ref{int-g4}) the exponent $m\geq 0$ always, and
these must not be analytically continued into the region of negative
values, whereas the exponents $i,j$ do follow the usual analytic
continuation process to get the final result for the integrals.

From our previous work \cite{tensor} on NDIM approach to tensorial
integrals, we know that the best solution for such kind of integrals
is a truncated \hf{} because it contains all the cases of interest in
the same formula: scalar, vector and arbitrary tensor rank.  The \hf{}
above is clearly truncated for positive integers $k$. This result,
among the five possible \hsr{} of such integral, is the only one that
is a truncated series for even and odd values of propagator exponent
$k$, since it assumes only positive values.

Finally, the result for the tensorial integral with three propagators,

\be \label{g4} 
g_4^{[{\rm AC}]}(i,j,k,m) =  \pi^{D/2} (-2)^m(\pp^2)^{\s'} (-i|-\p/2)
(\s'+\omega/2|j-\s')  (-j|\s') (-i-k-\p/2|-j-\omega/2)\;
_3F_2(\{4\}|1) ,
\ee
where $\s'=\s+m=i+j+k+m+D/2$. Note that the result (\ref{g4}) contains
the previous one, (\ref{g3}) in the particular case when $k=0$; it is
valid also when $i,j,k$ are negative and $m$ positive. The five
parameters $\{4\} \equiv \{a_4,b_4,c_4;e_4,f_4\}$ are given in the
table, and clearly the \hf{} is also truncated for even and odd
positive integers $m$.

The well-known \hf{} $_3F_2$ is defined by the series,
$$ _3F_2 \left.\left[ \matrix{ a, b, c \cr
                            e, f }\right| z\right] = \sum_{n=0}^\infty \frac{(a|n)(b|n) (c|n) 
}{ (e|n)(f|n)} \frac{z^n}{n!} , $$ 
so, when we refer to $_3F_2(...|1)$ we are meaning the above
series. For more details about \hf{}s the reader is referred to, e.g.,
\cite{luke}.

\begin{center}
\begin{tabular}{|c||c|c|} \hline 
 
{\rm Parameters} & ${}_3F_2(\{3\}|1)$ & ${}_3F_2(\{4\}|1)$ 
  \\ \hline\hline 
 
{\rm a} & $-k/2$ & $-m/2$  \\ \hline 
 
{\rm b} & $1/2-k/2$ & $1/2-m/2$ \\ \hline 
  
{\rm c} & $j+\omega/2 $ & $j+\omega/2 $ \\ \hline\hline 
  
{\rm e} & $1+i+j+D/2 $ & $1+i+j+k+D/2 $  \\ \hline  
 
{\rm f} & $1-i-k-D/2 $ & $1-i-k-m-D/2 $ \\ \hline\end{tabular} 
\end{center}
\begin{center} 
Parameters for hypergeometric functions in equations (\ref{g3}) and (\ref{g4}). 
\end{center}

\section{Two-loop Coulomb gauge integrals}

As far as we know, NDIM is the only approach where Feynman integrals
in different gauges, covariant and non-covariant alike, can be neatly
performed, without reference to any special prescription to handle
peculiar non-covariant singularities in the boson propagator. In the
usual covariant gauges several calculations were carried out, e.g.,
one-loop $n$-point function\cite{oleari}, scalar integrals for
photon-photon scattering in QED\cite{box} and genuine\cite{2-3loops}
two-loop three-point integrals. On the non-covariant side, we have
gotten an important original result: Light-cone integrals in the NDIM
context do not need the famous ML-prescription\cite{mandelstam} to
circumvent the gauge dependent singularities\cite{prescless}, as well
as avoiding other features which turn the calculation cumbersome --
such as using partial fractioning \cite{leib-nyeo} (a mandatory
feature there) and integration over components.

Coulomb gauge two-loop integrals can be treated within NDIM
methodology as well. Consider, for example,

\beq 
I_1(i,j,k,m) &= & \int d^D\! q\;d^D\!r\;\;
(q^2)^i(r^2)^j(p-r-q)^{2k}({\bf  r}^2)^m, \\
I_2(i,j,k,m) &= & \int d^D \! q\; d^D\! r\;\; (q^2)^i(q-r)^{2j}({\bf r}^2)^k 
({\bf p-q})^{2m}, \label{I2}
\eeq
which can be generated by,
\beq 
{\cal I}_1 &= & \int d^D\! q\;d^D\!r\;\; \exp{\left[ -\alpha q^2 - 
\beta
r^2 - \gamma (p-r-q)^2 -\theta {\bf r}^2 \right]} ,\\
{\cal I}_2 &= & \int d^D\! q\;d^D\!r\;\; \exp{\left[ -\alpha q^2 - 
\beta (q-r)^2 -\gamma {\bf r}^2 -\theta ({\bf p-q})^2 \right]}. 
\eeq

Following the usual steps of NDIM\cite{flying}, we get for the first
integral a $6\times 11$ matrix for the system of linear algebraic
equations to be solved
  
\be 
\left\{ \matrix{ X_{123} + Y_{12467} = i \cr
                     X_{13} + Y_{2368} = j \cr
                     X_{123} + Y_{13578} = k \cr
                     X_2 + Y_{45} = m \cr
                     X_{12} + Y_{12345} = -\omega/2 \cr
                     X_3 + Y_{678} = -\p/2 }\right. ,\label{sist-coul5}
\ee
which generates 462 ($6\times 6$) possible \hsr{} for the integral in
question. Of these, 216 have solutions in terms of hypergeometric series
whose variable is either $z= p_4^2/{\bf p}^2$ or $z^{-1}$.  Among these power
series the simplest ones are double series which we consider in more
detail. Of course, there are also series of the following form,

\be 
\sum_{a,b,c,e= 0}^\infty \frac{z^{a+b}}{a!b!} 
\frac{(z^{-1})^{c+e}}{c!e!} \frac{\G(...)}{\G(...)}, \qquad\quad 
\sum_{a,b,c,e= 0}^\infty \frac{z^{a}}{a!} 
\frac{(z^{-1})^{b+c+e}}{b!c!e!} \frac{\G(...)}{\G(...)}, 
\ee
which can only be convergent if $z= z^{-1}= 1$. Since this is a 
particular case, where $\pp^2= p_4^2$, we will not study it.

Let us consider the solution written in terms of double hypergeometric 
series,
\be 
I^{A,\;[{\rm AC}]}_1(i,j,k,m) =  \pi^D (p_4^2)^{\s''} P_A^{[{\rm AC}]} \; 
\sum_{n_1,n_2= 0}^\infty \frac{(-\s''|n_{12}) 
(-m|n_2) (m+\omega/2|n_1) (1-m-\s''-D/2|n_2) }{n_1!n_2! (1-i-k-m-D/2|n_2) 
(\omega/2|n_{12}) } \left(\frac{\pp^2}{p_4^2}\right)^{n_{12}} ,
\ee
where $\s''=\s'+D/2=i+j+k+m+D$ and $P_A^{[{\rm AC}]}$ is a product of
Pochhammer symbols,

\be 
P_A^{[{\rm AC}]} =  (-i|i+k+D/2)(-k|i+k+D/2)(\omega/2|m)(i+k+D|m)(-j|j-\s'')
(\s''+m+D/2|j-\s'')  ,
\ee
where the exponents of propagators $i,j,k$ must assume negative
values. There is another double series, in the other kinematical
region, where $|p_4^2/\pp^2| <1$, namely,

\be 
I^{B\;[{\rm AC}]}_1(i,j,k,m) =  \pi^D (\pp ^2)^{\s''} P_B^{[{\rm AC}]} \!   
\sum_{n_i= 0}^\infty \frac{(-\s''|n_2)  (-j-m-\omega/2|n_2-n_1)
(m+\omega/2|n_1) (j+m+D/2|n_1) }{n_1!n_2! (1-\s''-\omega/2|n_2-n_1) 
(\omega/2|n_1) } \left(\frac{p_4^2}{\pp^2}\right)^{n_{12}} ,
\ee 
where,

\be 
P_B^{[{\rm AC}]} = (-i|i-\s'')(-j|-m-\omega/2) (-k|j+k+m+D/2)
(\omega/2|m) (i+k+D|-k-D/2) (\p/2|k+\omega/2). 
\ee

The second integral is much simpler than the former, in that the
\hsr{} involved are all summable. Summing them is an easy task and the
result can be written in terms of gamma functions,

\beq 
\label{2loop-segunda} 
I^{[{\rm AC}]}_2(i,j,k,m) &= & \pi^D ({\bf p}^2)^{\s''}
(-i|-\p/2) (\s''+\omega/2|-2\s''-\omega/2) (-k|2k+\omega/2) (-m|2m+\omega/2)
(-j|i+ 2j+k+D) \\ 
&&\times (-i-j-k-D/2-\p/2|i+\p/2) (j+k+D-\p/2|-k-\omega/2).
\nonumber
\eeq

This is a 36-fold degenerate result, i.e., of the overall 56 ($5\times
5$) systems, 36 of them have non-trivial solutions, which after summed
(see e.g.  \cite{flying}) and analytically continued give
(\ref{2loop-segunda}). It is important to note that this result allows
negative values for $(i,j,k,m)$ and positive ones for $D$. For
negative $m$ see appendix.

\section{Conclusion}

Splitting the space-time dimension $D$ in the dimensional
regularization context into energy-sector and momentum sector, each
with a specific regularizing parameter, it was possible to Leibbrandt
{\it et al} to perform perturbative calculations in the Coulomb gauge
at one and two-loop level. However, the calculations are very involved
and they were able to present explicit results only for the divergent
parts of the integrals. On the other hand, using NDIM we showed here
that we can calculate complete results for the same integrals, and not
only that, they did not have to be carried out separately, In our
approach we can consider several of them at the same time, because we
leave the exponents of propagators arbitrary, the integrals being
either scalar or tensorial.

\acknowledgments{ AGMS gratefully acknowledges FAPESP (Funda\c 
c\~ao de Amparo \`a Pesquisa do Estado de S\~ao Paulo, Brasil) for  financial
support. } 
\vfill\eject

\appendix

\section{Special Cases: Extracting Poles}

To make things a little more illuminating we consider in this appendix
some technical issues relevant to the results for sample special cases.

\subsection{ One-loop.}

Let us consider for instance the particular case where the exponents
of propagators in the integral (\ref{int-g4}) are $i=j=-1, k=-2, m=2$. The
result for this integral is obtained from eq.(\ref{g4}),

\be 
g_4^{[{\rm AC}]}(-1,-1,-2,2) = 4\pi^{D/2}
(\pp^2)^{D/2-2}\frac{\G(1-\p/2)\G(\omega/2-1) \G(3-D/2)
\G(D/2-1)}{\G(3-\p/2)\G(D/2-2+\omega/2)} \;_3F_2 \left.\left( \matrix
{ -1,\; -1/2,\; \omega/2-1 \cr D/2-3,\; 2-D/2 }\right| 1\right), 
\ee
observing that now $\s'=D/2-2$. We proceed as usual in dimensional
regularization, taking $D=4-2\epsilon$, $\omega=3-\epsilon$,
$\p=1-\epsilon$ and Taylor expanding around $\epsilon=0$, to get

\be 
g_4^{[{\rm AC}]}(-1,-1,-2,2) = 4\pi^{2-\epsilon}(\pp^2)^{-\epsilon}
\left[ \frac{8}{3} +\left( -\frac{8\gamma_E}{3} + \frac{64}{9} -
\frac{16}{3}\ln{2}\right)\epsilon + {\cal O}(\epsilon^2)
\right]\;_3F_2 \left.\left( \matrix{ -1,\; -1/2,\; \omega/2-1 \cr
D/2-3,\; 2-D/2 }\right| 1\right), \label{gama1}
\ee

where $\gamma_E$ is the Euler's constant.

Now we turn to the theory of \hf{}s\cite{luke}. When a numerator
parameter is a negative integer the series is a truncated one. This is
exactly our case, and the \hf{} above has only two terms,

\be  
 _3F_2 \left.\left( \matrix{ -1,\; -1/2,\; \omega/2-1 \cr D/2-3,\;
                                    2-D/2 }\right| 1\right) = 1 +
                                    \frac{(-1|1)(-1/2|1)
                                    (\omega/2-1|1)}{1!(D/2-3|1)(2-D/2|1)}
                                    = 1 -
                                    \frac{(1-\epsilon)}{4\epsilon(1+\epsilon)},
\label{fator1}\ee
which substituted into (\ref{gama1}) yields,
\be 
g_4^{[{\rm AC}]}(-1,-1,-2,2) = 4\pi^{2-\epsilon}(\pp^2)^{-\epsilon}
\left( -\frac{2}{3\epsilon} + \frac{20}{9} + \frac{2\gamma_E}{3} +
\frac{4\ln{2}}{3}\right). 
\ee

We remember that the original integral was contracted with $4p_ip_j$,
so rewriting $4(\pp^2)^{-\epsilon} = 4(\pp^2)^{2-\epsilon-2} =
4(p_ip^ip_jp^j)(\pp^2)^{-2-\epsilon}$, we obtain the final result,

\be 
g_4^{[{\rm AC}]}(-1,-1,-2,2) =
\pi^{2-\epsilon}p_ip_j(\pp^2)^{-2-\epsilon} \left(
-\frac{2}{3\epsilon} + \frac{20}{9} + \frac{2\gamma_E}{3} +
\frac{4\ln{2}}{3}\right). 
\ee

\subsection{Two-loops.}

Let us consider two particular cases for integral (\ref{I2}): The
first one where $i=k=-2, j=-1, m=1$ and a second one where $i=k=-2,
j=m=-1$.

Observe that in the first case there is one exponent, $m$, which is
positive, so we must not analytic continue it (see for instance our
previous papers\cite{prescless,tensor}). The related Pochhammer symbol
$(-m|2m+\omega/2)$ was generated by,
$$ 
\frac{\G(1+m)}{\G(1-m-\omega/2)} = \frac{1}{\G(1+m|-2m-\omega/2)} =^{^{\!\!\!\!\!\!\! AC}} 
(-1)^{2m+\omega/2} (-m|2m+\omega/2). $$

However it must not be analytically continued since we are interested
in the special case where $m$ is positive ($m=1$). So the result for
(\ref{I2}), which allows one to take $m$ positive, reads now,

\beq  
I^{[{\rm AC}]}_2(i,j,k,m) &= & \pi^D ({\bf p}^2)^{\s''}
(-i|-\p/2) (\s''+\omega/2|-2\s''-\omega/2) (-k|2k+\omega/2)(-j|i+ 2j+k+D) \\ 
&&\times (-i-j-k-D/2-\p/2|i+\p/2) (j+k+D-\p/2|-k-\omega/2) \frac{\G(1+m)}{\G(1-m-\omega/2)} 
\nonumber\\
&&\times (-1)^{2m+\omega/2}.\nonumber
\eeq

Now it is easy to expand (we use the MAPLE V software) around
$D=4-2\epsilon$, with $\s''=D-4$, to obtain the poles, double and
simple ones, plus finite part
 
\beq  
I^{[{\rm AC}]}_2(-2,-1,-2,1) &=& -i^{1-\epsilon}\pi^{4-2\epsilon} ({\bf
p}^2)^{-2\epsilon} \left[ -\frac{1}{2\epsilon^2} +
\frac{7+12\ln{2}+6\gamma_E}{6\pi\epsilon} + \frac{-520 -112\gamma_E -
192\gamma_E\ln{2} +43\pi^2 -224\ln{2} }{48\pi} \right.\nonumber\\ && -
\left.\frac{48\gamma_E +192\ln^2{2}}{48\pi} +{\cal O}(\epsilon)\right].
\eeq

The second special case can be studied using (\ref{2loop-segunda})
directly since in this case all exponents $(i=k=-2, j=m=-1)$ are
negative, snd $(\s''=D-6)$. Using again MAPLE V software to expand
around $\epsilon=0$, we get a simple pole plus finite part,

\beq  
I^{[{\rm AC}]}_2(-2,-1,-2,-1) &=& \pi^{4-2\epsilon} ({\bf
p}^2)^{-2-2\epsilon} \left( \frac{1}{3\epsilon} + \frac{1}{3} -
\frac{2\gamma_E}{3} - \frac{4\ln{2}}{3}\right). 
\eeq

\end{document}